\documentclass[10pt,conference]{IEEEtran}
\usepackage{amssymb,amsmath,amscd}
\usepackage{cite,manfnt}
\usepackage{paralist}

\usepackage{url}

\newcommand{\vbf}[1]{\mathbf{#1}}

\newcommand{\F}{\mathbf{F}}

\newcommand{\Z}{\mathbf{Z}}

\newcommand{\mbf}{\mathbf}
\newcommand{\wt}{{\rm {wt}}}
\newcommand{\ord}{{\rm {ord}}}
\newcommand{\hdual}{{\bot_h}}
\newcommand{\scal}[2]{\langle #1|#2\rangle}
\newcommand{\scalh}[2]{\langle #1|#2\rangle_h}

\newcommand{\floor}[1]{\left\lfloor {#1}\right\rfloor}
\newcommand{\ceil}[1]{\left\lceil {#1}\right\rceil}

\newcommand{\nix}[1]{}

\def\H{\widetilde{H}}

\DeclareMathOperator{\coeff}{coeff} 
\DeclareMathOperator{\rk}{rk} 
 
\DeclareMathOperator{\ex}{ex}

\DeclareMathOperator{\im}{im}
\DeclareMathOperator{\kernel}{ker}

\newtheorem{theorem}{Theorem}
\newtheorem{lemma}[theorem]{Lemma}

\newtheorem{proposition}[theorem]{Proposition}

\begin{document}

\title{Quantum Convolutional BCH Codes}

\author{
\authorblockN{Salah A. Aly$^1$, Markus Grassl$^2$, Andreas Klappenecker$^1$,
Martin R\"otteler$^3$, Pradeep Kiran Sarvepalli$^1$}
\authorblockA{$^1$Department  of Computer Science, Texas A\&M University,
College Station, TX 77843-3112, USA \\
$^2$Institut f\"ur Algorithmen und Kognitive Systeme,
Universit\"at Karlsruhe (TH), D-76128 Karlsruhe, Germany\\
$^3$NEC Laboratories America, Inc., 4 Independence Way, Suite 200,
Princeton, NJ 08540, USA}
}
\maketitle

\begin{abstract}
  Quantum convolutional codes can be used to protect a sequence of
  qubits of arbitrary length against decoherence.  We introduce two new
  families of quantum convolutional codes. Our construction is based on
  an algebraic method which allows to construct classical
  convolutional codes from block codes,  in particular 
  BCH codes.  These codes have the property that they contain their
  Euclidean, respectively Hermitian, dual codes. Hence, they can be
  used to define quantum convolutional codes by the stabilizer code
  construction. We compute BCH-like bounds on the free distances which
  can be controlled as in the case of block codes, and establish that
  the codes have non-catastrophic encoders.
\end{abstract}

\section{Introduction}
Quantum convolutional codes provide an alternative to quantum block codes to protect
quantum information for reliable quantum communication. Ollivier and Tillich
launched the stabilizer framework for quantum convolutional codes~\cite{ollivier04}.
Using this stabilizer framework Forney {\em et al.}  constructed rate $(n-2)/n $
quantum convolutional codes \cite{forney05b}. Recently, two of us constructed
quantum convolutional codes from product codes \cite{grassl05} and 
derived an algorithm to construct non-catastrophic encoders and
encoder inverses~\cite{grassl06}. In~\cite{aly07b}, a generalized
Singleton bound for a class of quantum convolutional codes has been
established, together with a family of codes based on generalized
Reed-Solomon codes meeting this bound.  \nix{However, many questions remain
unanswered and there is a scarcity of good families of quantum
convolutional codes.}

Unit memory convolutional codes are an important class of codes that
appeared in a paper by Lee~\cite{lee76}. He also showed that these
codes have large free distance $d_f$ among other codes (multi-memory)
with the same rate. \nix{Upper and lower bounds on the free distance
of unit memory codes were derived by Thommesen and
Justesen~\cite{thommesen83}.} Convolutional codes are often designed
heuristically. However, classes of unit memory codes were constructed
algebraically by Piret based on Reed-Solomon codes~\cite{piret88} and
by Hole based on BCH codes~\cite{hole00}. 
%old Recently, subclasses of cyclic and doubly-cyclic convolutional codes
%old and their properties are described where the units of the skew
%old polynomial ring are used~\cite{gluesing04b}.
In a recent paper, doubly-cyclic convolutional codes are investigated
which include codes derived from Reed-Solomon and BCH codes
\cite{gluesing04b}.  These codes are related, but not identical to the
codes defined in this paper.

The main results of this paper are:
\begin{inparaenum}[(a)]
\item a method to construct convolutional codes from block codes
\item a new class of convolutional stabilizer codes based on BCH codes.
\end{inparaenum}
These codes have non-catastrophic dual encoders making it possible to
derive non-catastrophic encoders for the quantum convolutional codes.

\section{Background}
\subsection{Convolutional Codes}
We briefly recall the basic facts about classical convolutional codes relevant for
our discussion. Let $\F_q$ be a finite field with $q$ elements. A
\textit{convolutional code} $C$ of length $n$ and dimension $k$ over $\F_q$ is a
free module of rank $k$ that is a direct summand of $\F_q[D]^n$.  A matrix $G$ in
$\F_q[D]^{k\times n}$ such that $C=\im G=\{\vbf{u}G \mid\vbf{u}\in \F_q[D]^k\}$ is
called a \textit{basic generator matrix} of $C$, and a matrix $H\in
\F_q[D]^{(n-k)\times n}$ such that $C=\ker H^t =\{\vbf{v} \mid\vbf{v}\in
\F_q[D]^n,\vbf{v}H^t=0\}$ is called a \textit{basic parity check matrix} of $C$.

The existence of a convolutional code $C$ is equivalent to
the existence of four matrices $G \in \F_q[D]^{k\times n}$,
$H\in \F_q[D]^{(n-k)\times n}$, $K \in \F_q[D]^{n\times k}$,
and $L\,\in \F_q[D]^{n\times (n-k)}$
such that $C=\im G = \kernel H^t$, $GK=1_{\F_q[D]^k}$, and
$L^tH^t=1_{\F_q[D]^{n-k}}=HL$.

Let $\nu_i$ denote the maximum of the degrees among the polynomials in the $i$th row
of a basic generator matrix $G$, and let the \textit{memory} $m$ be the maximal
value of $\nu_i$.  A basic generator matrix of a convolutional code $C$ is called
\textit{reduced} if the {\em overall constraint length} $\nu=\nu_1+\cdots + \nu_k$
has the smallest value among all basic generator matrices of $C$.    It is often
convenient to express the generator matrix as
 $G=G_0+G_1 D+\cdots+G_m D^m$, where $G_i\in \F_q^{k\times n}$.

Let $\F_q((D))$ be the field of Laurent series consisting of elements of the form
$v(D)=\sum_{i} v_i D^i$ for $v_i\in \F_q$ and $v_i=0$ for $i\leq r$  for some $r\in
\Z$.  We associate with a convolutional code $C$ another module $C^\infty = \{
\mbf{u}(D)G \mid \mbf{u}(D)\in \F_q((D))^k\}$, The entries of a generator matrix $G$
of $C^\infty$ might be rational functions. Let $\mbf{v}(D) =(v_1(D),\ldots,
v_n(D))\in \F_q((D))^n$ where $v_i(D) =\sum_{j}v_{ij}D^j$. Then we can identify
$\mbf{v}(D)$ with an element in $\F_q^n((D))$ as $\sum_{j}\vbf{v}_j D^j$, where
$\vbf{v}_j=(v_{1j}, \ldots, v_{nj}) \in \F_q^n$. We define the weight of
$\mbf{v}(D)$ as $\wt(\mbf{v}(D))=\sum_{i\in \Z}\wt(\vbf{v}_i)$. A generator matrix
$G$ is called \textit{catastrophic} if there exists a $\mbf{u}(D)\in \F_q((D))^k$ of
infinite Hamming weight such that $\mbf{u}(D)G\in C^\infty$ has finite Hamming
weight.  The free distance $d_f$ of $C$ is defined as
\begin{eqnarray}
d_f =\min \{ \wt (\mbf{v}(D))\mid \mbf{v}(D) \in C,\mbf{v}(D)\ne 0  \}.
\end{eqnarray}
A rate $k/n$ convolutional code with memory $m$, overall constraint length $\nu$,
and free distance $d_f$ is denoted by $(n,k,\nu; m, d_f)_q$. Sometimes a shorter
notation $(n,k,\nu)_q$ is also used.

The \textit{Euclidean inner product} of two $n$-tuples $\mbf{u}(D)=\sum_i\vbf{u}_i
D^i$ and $\mbf{v}(D)=\sum_j\vbf{v}_i D^j$ in $\F_q[D]^n$ is defined as $
\scal{\mbf{u}(D)}{\mbf{v}(D)} = \sum_{i} \vbf{u}_i\cdot \vbf{v}_i$. The Euclidean
dual of a convolutional code $C$ is denoted by $C^\perp=\{ \mbf{u}(D)\in
\F_q[D]^n\mid \scal{\mbf{u}(D)}{\mbf{v}(D)}=0 \text{ for all } \mbf{v}(D)\in C\}$.
Note that $H(D)$, the parity check matrix of $C$, \textit{does not} generate the
Euclidean dual in general. Instead, one has to reverse the order of the coefficients
of the polynomials in $H(D)$, {\em i.e.} consider the matrix $D^{m^\perp} H(1/D)$,
where $m^\perp$ is the memory of the code generated by $H(D)$. For codes over
$\F_{q^2}$, we define the Hermitian inner product as $\scalh{\mbf{u}(D)}{\mbf{v}(D)}
= \sum_{i} \vbf{u}_i\cdot \vbf{v}_i^q $, where $\vbf{u}_i,\vbf{v}_i\in \F_{q^2}^n$
and $\vbf{v}_i^q=(v_{1i}^q,\ldots,v_{ni}^q)$.  The Hermitian dual of $C$ is then
$C^\hdual=\{ \mbf{u}(D)\in \F_{q^2}[D]^n\mid \scalh{\mbf{u}(D)}{\mbf{v}(D)}=0 \text{
for all } \mbf{v}(D)\in C\}$.

\subsection{Quantum Convolutional Codes}

We briefly describe the stabilizer framework for quantum convolutional codes, see
also \cite{aly07b,grassl07,ollivier04}. The stabilizer is
given by a matrix
\begin{equation}\label{stab-mat}
S(D)=(X(D)|Z(D))
\in\F_q[D]^{(n-k)\times 2n}.
\end{equation}
which satisfies the symplectic orthogonality condition $0 = X(D) Z(1/D)^t
- Z(D) X(1/D)^t$.  Let ${\cal C}$ be a quantum convolutional code
defined by a stabilizer matrix as in eq.~(\ref{stab-mat}). Then $n$ is
called the frame size, $k$ the number of logical qudits per frame, and
$k/n$ the rate of ${\cal C}$. It can be used to encode a sequence of
blocks with $k$ qudits in each block (that is, each element in the
sequence consists of $k$ quantum systems each of which is
$q$-dimensional) into a sequence of blocks with $n$ qudits.

 The memory of the quantum convolutional code is defined as $m = \max_{1 \leq i
\leq n-k,1 \leq j \leq n}(\max(\deg X_{ij}(D),\deg Z_{ij}(D)))$. We use the notation
$[(n,k,m)]_q$ to denote a quantum convolutional code with the above parameters. We
can identify $S(D)$ with the generator matrix of a self-orthogonal classical
convolutional code over $\F_q$ or $\F_{q^2}$, which gives us a means to construct
convolutional stabilizer codes. Analogous to the classical codes we can define the
free distance, $d_f$ and the degree $\nu$, prompting an extended notation
$[(n,k,m;\nu,d_f)]_q$. All the parameters of the quantum convolutional code can be
related to the associated classical code as the following propositions will show.
For proof and further details see \cite{aly07b}\footnote{A small difference exists
between the notion of memory defined here and the one used in
\cite{aly07b}.}.

\begin{proposition}\label{pr:css}
  Let $(n,(n-k)/2,\nu;m)_q$ be a convolutional code such that $C \leq
  C^\perp$, where the dimension of $C^\perp$ is given by $(n+k)/2$.
  Then an $[(n,k,m;\nu,d_f)]_q$ convolutional stabilizer code exists
  whose free distance is given by $d_f=\wt(C^\perp \backslash C)$, which
  is said to be pure if $d_f = \wt(C^\perp)$.
\end{proposition}

\begin{proposition}\label{pr:c2qHerm}
Let $C$ be an $(n,(n-k)/2,\nu;m)_{q^2}$ convolutional code such that
$C\subseteq C^\hdual$.  Then there exists an $[(n,k,m;\nu,d_f)]_q$
convolutional stabilizer code, where $d_f=\wt(C^\hdual\setminus C)$.
\end{proposition}

\section{A Construction of Convolutional Codes}
In this section, we give a method to construct convolutional codes from block codes.
This generalizes an earlier construction by Piret \cite{piret88b} to construct
convolutional codes from block codes. One benefit of this method is that we can
easily bound the free distance using the techniques for block codes. Another benefit
is that we can derive non-catastrophic encoders.

\subsection{Convolutional Codes from Block Codes}
Given an $[n,k,d]_q$ block code with parity check matrix $H$, it is possible to
split the matrix $H$ into $m+1$ disjoint submatrices $H_i$, each of which has $n$
columns, such that
\begin{eqnarray}
H=\left[\begin{array}{c} H_0\\H_1\\ \vdots\\ H_m
\end{array}\right]. \label{eq:splitH}
\end{eqnarray}
Then we can form the polynomial matrix
\begin{eqnarray}
G(D)=\H_0+\H_1 D+\H_2 D^2+\ldots+\H_m D^m,\label{eq:ccH}
\end{eqnarray}
where the number of rows of $G(D)$ equals the maximal number $\kappa$
of rows among the matrices $H_i$.  The matrices $\H_i$ are obtained
from the matrices $H_i$ by adding zero-rows 
at the bottom such that the matrix $\H_i$ has $\kappa$ rows in
total. Then $G(D)$ generates a convolutional code. The fact that the
$H_i$ come from a common block code allows us to characterize the
parameters of the convolutional code and its dual using the techniques
of block codes. Our first result concerns a non-catastrophic encoder
for the code generated by $G(D)$.

\begin{theorem}\label{th:noncataDualEnc}
  Let $C\subseteq \F_q^n$ be an $[n,k,d]_q$ linear code with parity
  check matrix $H \in \F_q^{(n-k)\times n}$. Assume that $H$ is
  partitioned into submatrices $H_0,H_1,\ldots,H_m$ as in
  eq.~(\ref{eq:splitH}) such that $\kappa = \rk H_0$ and $\rk
  H_i\le \kappa$ for $1\le i\le m$. Define the polynomial matrix
  $G(D)$ as in eq.~(\ref{eq:ccH}). Then we have:
\begin{compactenum}[(a)]
\item \label{lm:CCbasic}
The matrix $G(D)$ is a reduced basic generator matrix.

\item \label{lm:CCdual} If the code $C$ contains its Euclidean dual
$C^\bot$, respectively its Hermitian dual $C^\hdual$, then the
convolutional code $V=\{\mbf{v}(D) = \mbf{u}(D) G(D)\mid
\mbf{u}(D)\in \F_q^{n-k}[D]\}$ is contained in its dual $V^\perp$,
respectively its Hermitian dual $V^\hdual$.

\item \label{lm:CCdist}
Let $d_f$ and $d_f^\perp$ respectively denote the free distances of $V$ and
$V^\perp$. Let $d_i$ be the minimum distance of the code $C_i=\{\vbf{v}\in
\F_q^n\mid\vbf{v}\H_i^t =0\}$, and let $d^\perp$ denote the minimum distance of
$C^\perp$. Then the free distances are  bounded by $\min \{d_0+d_m,d \}\leq
d_f^\perp\leq d$ and $d_f \geq d^\perp$.
\end{compactenum}
\end{theorem}

\begin{proof}
  To prove the claim (a), it suffices to show that (i) $G(0)$ has full
  rank $\kappa$, (ii) $(\coeff(G(D)_{ij},D^{\nu_i}))_{1\le i\le
    \kappa, 1\le j\le n}$, has full rank $\kappa$,
where for $f(D)=\sum_{i\geq 0} a_i D^i$ we define $\coeff(f(D),D^i)=a_i$, and (iii)
$G(D)$ is non-catastrophic; cf.~\cite[Theorem 2.16 and Theorem 2.24]{piret88}.

By definition, $G(0)=\H_0$ has rank $\kappa$, so (i) is
satisfied. Condition (ii) is satisfied, since the rows of $H$ are
linearly independent; thus, the rows of the highest degree coefficient
matrix are independent as well.

It remains to prove (iii). Seeking a contradiction, we assume that the generator
matrix $G(D)$ is catastrophic. Then there exists an input sequence
$\mathbf{u}(D)=\sum_i\vbf{u}_iD^i\in \F_q((D))^\kappa$ with infinite Hamming weight
that is mapped to an output sequence
$\mathbf{v}(D)=\mathbf{u}(D)G=\sum_i\vbf{v}_iD^i\in\F_q((D))^n$ with finite Hamming
weight, {\em i.e.} $\vbf{v}_i=0$ for all $i\ge i_0$. We have
\begin{equation}\label{eq:encoding}
\vbf{v}_{i+m} = \vbf{u}_{i+m} \H_0 + \vbf{u}_{i+m-1}\H_1+\ldots+\vbf{u}_i\H_m,
\end{equation}
where $\vbf{v}_{i+m}\in\F_q^n$ and $\vbf{u}_j\in\F_q^\kappa$.  By construction,
the vector spaces generated by the rows of the matrices $H_i$
intersect trivially. Hence $\vbf{v}_i=0$ for $i\ge i_0$ implies that
$\vbf{u}_{i-j}\H_j=0$ for $j=0,\ldots,m$.  The matrix $\H_0$ has full
rank. This implies that $\vbf{u}_i=0$ for $i\ge i_0$, contradicting the fact
that $\mbf{u}(D)$ has infinite Hamming weight; thus, the claim (a) holds.

To prove the claim (b), let $\mbf{v}(D)=\sum_i\vbf{v}_iD^i$,
$\mbf{w}(D)=\sum_i\vbf{w}_i D^i$ be any two codewords in $V\subseteq \F_q^n[D]$.
Then from eq.~(\ref{eq:encoding}), we see that $\vbf{v}_i$ and $\vbf{w}_j$ are in
the rowspan of $H$ {\em i.e.} they are elements of $C^\perp$, for any $i,j\in \Z$.
Since $C^\perp \subseteq C = (C^\perp)^\perp$, it follows that $\vbf{v}_i\cdot
\vbf{w}_j=0 $, for any $i,j \in \Z$ which implies that
$\scal{\mbf{v}(D)}{\mbf{w}(D)} =\sum_{i\in\Z}\vbf{v}_i\cdot \vbf{w}_i =0$.  Hence
$V\subseteq V^\perp$.  Similarly, we can show that if $C^\hdual \subseteq C$, then
$V \subseteq V^\hdual$.

For the claim (c), without loss of generality assume that the codeword $\mbf{c}(D)
=\sum_{i=0}^\ell\vbf{c}_iD^i$ is in $V^\perp$, with $\vbf{c}_0\ne 0
\ne\vbf{c}_\ell$.

It follows that $\scal{D^i\mbf{c}(D)}{D^l G_j(D)}=0$ for $i,l\ge 0$, where $G_j(D)$
denotes the $j$th row of $G(D)$.  In particular we have $\vbf{c}_0\H_m^t=0$ and
$\vbf{c}_\ell\H_0^t=0$.  It follows that $\vbf{c}_0\in C_m$ and $\vbf{c}_\ell\in
C_0$.  If $\ell>0$, then $\wt(\vbf{c}_0)\geq d_m$ and $\wt(\vbf{c}_\ell)\geq d_0$
implying $\wt(\mbf{c}(D))\geq d_0+d_m$. If $\ell=0$, then
$\scal{D^i\vbf{c}_0}{G_j(D)}=0$ implies $\vbf{c}_0\H_i^t=0$ for $0\leq i\leq m$,
whence $\vbf{c}_0 H^t=0$ and $\vbf{c}_0\in C$, implying that $\wt(\vbf{c}_0)\geq d$.
It follows that $\wt(\mbf{c}(D))\geq \min \{ d_0+d_m, d\}$, giving the lower bound
on $d_f^\perp$.

For the upper bound note that if $\vbf{c}_0$ is a codeword of $C$, then
$\vbf{c}_0H_i^t=0$.  From $\vbf{c}_0$ we can construct a codeword $\mbf{c}(D)$ by
padding with zeros.  Now, $\scal{D^i\mbf{c}(D)}{D^lG_j(D)}=0$ and hence
$\mbf{c}(D)\in V^\perp$. Since $\wt(\mbf{c}(D))=\wt(\vbf{c}_0)$ we obtain that
$d_f^\perp\leq d$.

Finally, let $\mbf{c}(D)=\sum_i\vbf{c}_iD^i$ be a non-zero codeword in $V$. We saw
earlier in the proof of (b) that every $\vbf{c}_i$ is in $C^\perp$. Thus $d_f\geq
\min\{\wt(\vbf{c}_i)\mid\vbf{c}_i\ne 0\} \geq d^\perp$.
\end{proof}

A special case of our claim (a) has been established by a different method
in~\cite[Proposition 1]{hole00}.

\section{Convolutional BCH Codes}
One of the attractive features of BCH codes is that they allow us to
design codes with desired distance. There have been prior approaches
to construct convolutional BCH codes, see \cite{hole00,rosenthal99},
and most notably \cite{gluesing04b}, where one can control the free
distance of the convolutional code.  Here we focus on codes with unit
memory.  Our codes have better distance parameters as compared to
Hole's construction \cite{hole00} and are easier to construct compared
to \cite{rosenthal99}.

\subsection{Unit Memory Convolutional BCH Codes}
Let $\F_q$ be a finite field with $q$ elements, $n$ be a positive
integer such that $\gcd(n,q)=1$. Let $\alpha$ be a primitive $n$th
root of unity.  A BCH code $C$ of designed distance $\delta$ and
length $n$ is a cyclic code with generator polynomial $g(x)$ in
$\F_q[x]/\langle x^n-1\rangle$ whose defining set is given by
$Z=C_b\cup C_{b+1}\cup \cdots \cup C_{b+\delta-2}$, where
$C_x=\{xq^i\bmod n \mid i\in \Z, i\ge 0 \}$. Let
\[\arraycolsep0.9\arraycolsep
H_{\delta,b} =\left[ \begin{array}{ccccc}
1 &\alpha^b &\alpha^{2b} &\cdots &\alpha^{b(n-1)}\\
1 &\alpha^{b+1} &\alpha^{2(b+1)} &\cdots &\alpha^{(b+1)(n-1)}\\
\vdots& \vdots &\vdots &\ddots &\vdots\\
1 &\alpha^{(b+\delta-2)} &\alpha^{2(b+\delta-2)} &\cdots &\alpha^{(b+\delta-2)(n-1)}
\end{array}\right].
\]
Then $C=\{\vbf{v}\in \F_q^n\mid\vbf{v}H_{\delta,b}^t=0\}$. If $r=\ord_n(q)$, then a
parity check matrix $H$ for $C$ is given by writing every entry in the matrix
$H_{\delta,b}$ as a column vector over some $\F_q$-basis of $\F_{q^r}$, and removing
any dependent rows.  Let $B=\{b_1,\dots,b_r\}$ denote a basis of $\F_{q^r}$ over
$\F_q$. Suppose that $\vbf{w}=(w_1,\dots,w_n)$ is a vector in $\F_{q^r}^n$, then we
can write $w_j = w_{j,1}b_1+ \cdots + w_{j,r}b_r$ for $1\le j\le n$. Let
$\vbf{w}^{(i)}=(w_{1,i},\dots, w_{n,i})$ be vectors in $\F_q^n$ with $1\le i\le r$,
For a vector $\vbf{v}$ in $\F_q^n$, we have $\vbf{v}\cdot\vbf{w}=0$ if and only if
$\vbf{v}\cdot\vbf{w}^{(i)}=0$ for all $1\le i\le r$.

For a matrix $M$ over $\F_{q^r}$, let $\ex_B(M)$ denote the matrix that is obtained
by expanding each row into $r$ rows over $\F_q$ with respect to the basis $B$, and
deleting all but the first rows that generate the rowspan of the expanded matrix.
Then $H=\ex_B(H_{\delta,b})$.

It is well known that the minimum distance of a BCH code is greater than or equal to
its designed distance $\delta$, which is very useful in constructing
codes\cite{huffman03}.  Before we can construct convolutional BCH codes we need the
following result on the distance of cyclic codes.

\begin{lemma}\label{lm:hartman}
Let $\gcd(n,q)=1$ and $2\leq \alpha\leq \beta <n$. Let $C\subseteq
\F_q^n$ be a cyclic code with defining set

\begin{equation}\label{eq:defining_set}
Z=\{z\mid z\in C_x, \alpha\le x\le\beta, x\not\equiv 0\bmod q\}.
\end{equation}
 The minimum distance $\Delta(\alpha,\beta)$ of $C$ is lower bounded as

\[
\Delta(\alpha,\beta) \geq
\begin{cases}
q+\floor{(\beta-\alpha+3)/q}-2, & \text{if $\beta-\alpha \geq 2q-3$;}\\
\floor{(\beta-\alpha+3)/2}, & \text{otherwise.}
\end{cases}
\]
\end{lemma}
\medskip

\begin{proof}
Our goal is to bound the distance of $C$ using the Hartmann-Tzeng bound (for
instance, see \cite{huffman03}). Suppose that there exists $a$ such that
$A=\{z,z+1,\ldots, z+a-2 \} \subseteq Z$. Suppose further, that there exists $b$,
where $\gcd(b,q)<a$ and $A+jb =\{ z+jb,z+1+jb,\ldots, z+a-2+jb\} \subseteq Z$ for
all $0\leq j\leq s$. Then by \cite[Theorem~4.5.6]{huffman03}, the minimum distance
of $C$ is $\Delta(\alpha,\beta) \geq a+s$.

We choose $b=q$, so that $\gcd(n,q)=1<a$ is satisfied for any $a>1$.
Next we choose $A\subseteq Z$ such that $|A|=q-1$ and $A+jb\subseteq Z$ for $0\leq j\leq s$,
with $s$ as large as possible. Now two cases can arise.
If $\beta-\alpha+1 < 2q-2$, then there {\em may not} always exist a set $A$ such
that $|A|=q-1$. In this case we relax the constraint that
$|A|=q-1$ and choose $A$ as the set of maximum number of consecutive elements.
Then $|A|=a-1 \geq \floor{(\beta-\alpha+1)/2}$ and $s\geq 0$ giving
the distance $\Delta(\alpha,\beta)\geq \floor{(\beta-\alpha+1)/2}+1= \floor{\beta-\alpha+3)/2}$.

If $(\beta-\alpha+1) \geq 2q-2$, then we can always choose a set
$A\subseteq \{z \mid \alpha \leq z\leq \alpha+2q-3, z\not\equiv 0\bmod q\}$ such that $|A|=q-1$.
As we want to make $s$ as large as possible, the worst case arises
when $A=\{\alpha+q-1,\ldots,\alpha+2q-3\}$.  Since $A+jb\subseteq Z$
holds for $0\leq j\leq s$, it follows $\alpha+2q-3+sq\leq \beta$. Thus
$s \leq \floor{(\beta-\alpha+3)/q}-2$.  Thus the distance
$\Delta(\alpha,\beta)\geq q+\floor{(\alpha-\beta+3)/q}-2$.
\end{proof}

\begin{theorem}[Convolutional BCH codes]\label{th:bchCC}
Let $n$ be a positive integer such that $\gcd(n,q)=1$, $r=\ord_n(q)$
and $2\leq 2\delta <\delta_{\max}$, where
$$\delta_{\max}=\left\lfloor\frac{n}{q^{r}-1} (q^{\lceil
r/2\rceil}-1-(q-2)[r \textup{ odd}])\right\rfloor.$$ Then there exists a unit memory
rate $k/n$ convolutional BCH code with free distance $d_f\geq
\delta+1+\Delta(\delta+1,2\delta)$ and $k=n-\kappa$, where
$\kappa=r\ceil{\delta(1-1/q)}$.  The free distance of the dual is $\geq
\delta_{\max}+1$.
\end{theorem}
\begin{proof}
Let $C\subseteq \F_q^n$ be a narrow-sense BCH code of designed
distance $2\delta+1$ and let $B$ a basis of $\F_{q^r}$ over $\F_q$.
Recall that a parity check matrix for $C$ is given
by $H=\ex_B(H_{2\delta+1,1})$.
Further, let $H_0=\ex_B(H_{\delta+1,1})$,  then from
\begin{eqnarray}
H_{2\delta+1,1}=\left[ \begin{array}{c}H_{\delta+1,1}\\
H_{\delta+1,\delta+1}\end{array}\right],\label{eq:bchH}
\end{eqnarray}
 it follows that 
$H=\left[H_0^t, H_1^t\right]^t$, where
$H_1$ is obtained from $\ex_B(H_{\delta+1,\delta+1})$ by
removing all rows common to $\ex_B(H_{\delta+1,1 })$. The code $C_0$
with parity check matrix $H_0=\ex_B(H_{\delta+1,1})$ coincides
with the narrow-sense BCH code of length $n$ and designed distance
$\delta+1$.

By \cite[Theorem~10]{aly07a}, we have $\dim C = n-
r\ceil{2\delta(1-1/q)}$ and $\dim C_0 = n-r\ceil{\delta(1-1/q)}$
which implies $\rk H= r\ceil{2\delta(1-1/q)}$, $\rk H_0 =
r\ceil{\delta(1-1/q)}$, and $\rk H_1 = \rk H - \rk H_0 =
r\ceil{2\delta(1-1/q)}-r\ceil{\delta(1-1/q)}$. For $x>0$, we have
$\ceil{x} \geq \ceil{2x} -\ceil{x}$; therefore, $\kappa= \rk H_0 \geq
\rk H_1$.

By Theorem~\ref{th:noncataDualEnc}(\ref{lm:CCbasic}), the matrix $H$ defines a reduced basic
generator matrix
\begin{eqnarray}
 G(D) = \H_0+D\H_1 \label{eq:bchHD}
\end{eqnarray}
of a convolutional code of dimension $\kappa$, while its dual which we refer to
as a convolutional BCH code is of dimension $n-\kappa$.

Now $H_1$ is the parity check matrix of a cyclic code, $C_1$ of the form given in
Lemma~\ref{lm:hartman}, {\em i.e.} the defining set of $C_1$ is $Z_1$ as defined in
(\ref{eq:defining_set}) with  $\alpha =\delta+1$ and $\beta=2\delta$. Since $H_1$ is
linearly independent of $H_0$ we have $x\not\equiv 0\bmod q$ in the definition of
$Z_1$.

By Theorem~\ref{th:noncataDualEnc}(\ref{lm:CCdist}), the free distance of the convolutional BCH code
is bounded as $\min\{d_0+d_1,d \}\leq d_f\leq d$.
By Lemma~\ref{lm:hartman}, $d_1\geq\Delta(\delta+1,2\delta) $ and by the
BCH bound $d_0\geq \delta+1$. Thus $d_f\geq \delta+1+\Delta(\delta+1,2\delta)$.
The dual free distance also follows from Theorem~\ref{th:noncataDualEnc}(\ref{lm:CCdist}) as
$d_f^\perp\geq d^\perp$. But $d^\perp \geq \delta_{\max}+1$ by
\cite[Lemma~12]{aly07a}.
\end{proof}

\section{Constructing Quantum Convolutional Codes}
Under some restrictions on the designed free distance, we can use convolutional
codes derived in the previous section to construct quantum convolutional codes.

\begin{theorem}\label{th:bchQccEuclid}
  Assume the same notation as in Theorem~\ref{th:bchCC}. Then there
  exists a quantum convolutional code ${\cal C}$ with parameters
  $[(n,n-2\kappa,1)]_q$, where $\kappa = r\ceil{\delta(1-1/q)}$.  For
  the free distance of ${\cal C}$ the bound $d_{f}\geq
  \delta+1+\Delta(\delta+1,2\delta)$ holds and it is pure to $d'\geq
  \delta_{\max}+1$.
\end{theorem}
\begin{proof}
We construct a unit memory $(n,n-\kappa)_q$ classical convolutional BCH code as per
Theorem~\ref{th:bchCC}. Its polynomial parity check matrix $G(D)$ is as given in
eq.~(\ref{eq:bchHD}). Using the notation as in the proof of Theorem~\ref{th:bchCC},
we see that the code contains its dual if $H$ is self-orthogonal. But given the
restrictions on the designed distance, we know from \cite[Theorem~3]{aly07a} that
the BCH block code defined by $H$ contains its dual. It follows from
Theorem~\ref{th:noncataDualEnc}(\ref{lm:CCdual}) that the convolutional BCH code
contains its dual.  From Proposition~\ref{pr:css} we can conclude that there exists
a convolutional code with the parameters $[(n,n-2\kappa,1)]_q$. By
Theorem~\ref{th:bchCC} the free distance of the dual is $d' \geq \delta_{\max}+1$,
also implying its purity.
\end{proof}

Another useful method to construct quantum codes makes use of codes over $\F_{q^2}$.

\begin{theorem}\label{lem:QCCHermitianBCH}
Let $2\leq 2\delta  < \floor{n(q^r-1)/(q^{2r}-1)} $, where and $r=\ord_n(q^2)$. Then
there exist quantum convolutional codes with parameters $ [(n,n-2\kappa,1)]_q$ and
free distance $d_{f} \geq \delta+1+\Delta(\delta+1,2\delta)$, where
$\kappa=r\ceil{\delta(1-1/q^2)}$.
\end{theorem}
\begin{proof}
By Theorem~\ref{th:bchCC} there exists an $(n,n-\kappa,1)_{q^2}$
convolutional BCH code with the polynomial parity check matrix as in
eq.~(\ref{eq:bchHD}). The parent BCH code has design distance
$2\delta+1$ and given the range of $\delta$, we know by
\cite[Theorem~14]{aly07a} that it contains its Hermitian dual. By
Theorem~\ref{th:noncataDualEnc}(\ref{lm:CCdual}), the convolutional
code also contains its Hermitian dual. By
Proposition~\ref{pr:c2qHerm}, we can conclude that there exists an
$[(n,n-2\kappa,1)]_q$ code with $d_f\geq
\delta+1+\Delta(\delta+1,2\delta)$.
\end{proof}
We conclude by noting that the convolutional codes in Theorems
\ref{th:bchQccEuclid} and \ref{lem:QCCHermitianBCH} have
non-catastrophic encoders and encoder inverses.  This follows directly
from the fact that $G(D)$ in eq.~(\ref{eq:bchHD}) is a basic generator
matrix (cf. \cite{grassl06,grassl07}).

\section*{Acknowledgment}
We would like to thank one of the referees for drawing our attention
to \cite{gluesing04b}.  This research was supported by NSF CAREER
award CCF~0347310, NSF grant CCF~0622201, and a Texas A\&M TITF
initiative.

\vspace*{0pt plus 0pt minus 6 pt}

% Generated by IEEEtranS.bst, version: 1.12 (2007/01/11)

\end{document}